\documentclass[aps,pra,superscriptaddress,showpacs,twocolumn]{revtex4-1}
\usepackage{bm}
\usepackage{amsmath}
\usepackage{float}
\usepackage{longtable}
\usepackage{dcolumn}
\newcolumntype{.}{D{x}{}{-1}}

\newcolumntype{w}[1]{D{.}{.}{#1}}

\newcommand{\Za}{{Z\alpha}}

\usepackage{graphicx}
\usepackage{xcolor}
\def\mynote1#1{{\color{blue}{\textsc{\footnotesize Question/Comment:} #1}}}
\def\mynote2#1{{\color{magenta}{#1}}}

\begin{document}

\title{Electron-correlation effects in the $\bm{g}$-factor of light Li-like ions}

\author{V. A. Yerokhin}
\affiliation{Center for Advanced Studies, Peter the Great St.~Petersburg Polytechnic University,
Polytekhnicheskaya 29, 195251 St.~Petersburg, Russia}

\author{K. Pachucki}
\affiliation{Faculty of Physics, University of Warsaw,
             Pasteura 5, 02-093 Warsaw, Poland}

\author{M. Puchalski}
\affiliation{Faculty of Physics, University of Warsaw,
             Pasteura 5, 02-093 Warsaw, Poland}
\affiliation{Faculty of Chemistry, Adam Mickiewicz University,
             Umultowska 89b, 61-614 Pozna{\'n}, Poland}

\author{Z. Harman}
\affiliation{Max~Planck~Institute for Nuclear Physics, Saupfercheckweg~1, D~69117 Heidelberg,
Germany}

\author{C.~H. Keitel}
\affiliation{Max~Planck~Institute for Nuclear Physics, Saupfercheckweg~1, D~69117 Heidelberg,
Germany}

\begin{abstract}

We investigate electron-correlation effects in the $g$-factor of the ground state of Li-like
ions. Our calculations are performed within the nonrelativistic quantum electrodynamics (NRQED)
expansion up to two leading orders in the fine-structure constant $\alpha$, $\alpha^2$ and
$\alpha^3$. The dependence of the NRQED results on the nuclear charge number $Z$ is studied and
the individual $1/Z$-expansion contributions are identified. Combining the obtained data with the
results of the all-order (in $\Za$) calculations performed within the $1/Z$ expansion, we derive
the unified theoretical predictions for the $g$-factor of light Li-like ions.

\end{abstract}

\maketitle

\section*{Introduction}

Measurements of the bound-electron $g$-factor in light H-like ions
\cite{sturm:11,sturm:13:Si,sturm:14} provide one of the best tests of the bound-state QED theory as
well as the most accurate determination of the electron mass \cite{mohr:16:codata}. Similar
experiments on Li-like ions \cite{wagner:13,sturm:16} probe the QED theory of the
electron-correlation effects. In future, a combination of $g$-factor experiments on Li-like and
H-like ions can be used as a new way to determine the fine-structure constant $\alpha$
\cite{yerokhin:16:gfact:prl}.

The QED effects in the $g$-factors of few-electron atoms can be systematically treated within the
two methods. The first method starts with the Dirac equation for the valence electron in a Coulomb
field of the nucleus and accounts for the radiative and electron-electron interaction effects by
perturbation theory. The expansion parameter for the electron-electron interaction is $1/Z$ (where
$Z$ is the nuclear charge number). This method accounts for all orders in the nuclear binding
strength parameter $\Za$ and thus is most effective for high-$Z$ atoms. Extensive QED calculations
of the $g$-factors of Li-like ions within the $1/Z$ expansion method were performed by Shabaev and
co-workers \cite{shabaev:02:li,glazov:04:pra,glazov:06:pla,glazov:10,volotka:14}.

The starting point of the second method is the Schr\"odinger equation that includes both the
electron-nucleus and the electron-electron Coulomb interactions. The radiative and relativistic
effects are accounted for by perturbation theory, with the expansion parameters $\alpha$ and $\Za$,
respectively. This method is often denoted as the Nonrelativistic Quantum Electrodynamics (NRQED)
approach, since the coefficients of the perturbation expansion can be derived systematically within
NRQED. In contrast to the first method, the NRQED treatment accounts for all orders in $1/Z$ but
expands in $\Za$ and thus is most effective for low-$Z$ atoms. Calculations of $g$-factors by this
method were carried out by Hegstrom~\cite{hegstrom:73} and, more recently, by
Yan~\cite{yan:01:prl,yan:02:jpb}.

The experiments on the $g$-factor of Li-like atoms have been so far performed in the intermediate
region of $Z$, where the two methods are complementary to each other. The optimal theoretical
treatment in this region of $Z$ can be achieved by combining them together. To this end, one would
need to identify (i) the individual $1/Z$-expansion terms in the NRQED calculations and (ii) the
individual $\Za$-expansion terms in the $1/Z$-expansion results. A combination of these results
would then provide a unified theory. The goal of the present investigation is to make the first
steps along this path.

In this  work we perform the NRQED calculation of the leading relativistic ($\sim\!\alpha^2$) and
the leading QED ($\sim\!\alpha^3$) corrections to the $g$-factor of the ground state of Li-like
ions, extending previous calculations by Yan to a larger region of $Z$ and improving the numerical
accuracy. We identify the individual $1/Z$ expansion terms of these corrections. In particular, we
obtain the higher-order electron-correlation contribution of the relative order $1/Z^3$ and higher,
thus removing one of the dominant sources of the uncertainty of theoretical predictions
\cite{volotka:14}.

\section{NRQED approach}

Within the NRQED approach, the bound-electron $g$-factor of a light atom is represented as an
expansion in powers of the fine-structure constant $\alpha$,
\begin{align}
g = g_e + \alpha^2\, g^{(2)} + \alpha^3\, g^{(3)} + \alpha^4\, g^{(4)} + \ldots\,,
\end{align}
where $g_e$ is the free-electron $g$-factor and $g^{(n)}$ are the binding corrections. The
expansion coefficients $ g^{(n)}$ can be further expanded in powers of the electron-to-proton mass
ratio $m/M$,
\begin{align}
g^{(n)} =  g^{(n)}_{\infty} + \frac{m}{M}\, g^{(n)}_M  + \ldots\,.
\end{align}

The interaction of a {\em free} nonrelativistic electron with a constant external magnetic field
$\vec B$ is described by the Hamiltonian
\begin{eqnarray}
H &=& \mu_B\,(1+\kappa)\,\vec\sigma\cdot\vec B\nonumber \\
   &=& \mu_B\,(1+\kappa)\,2\,\vec s\cdot\vec B
   \equiv \mu_B\,g_e\,\vec s\cdot\vec B\,,
\end{eqnarray}
where $\mu_B = -e/(2\,m)$ is the Bohr magneton, $\vec \sigma$ is the vector of Pauli matrices,
$\vec s$ is the electron spin operator, $\kappa$ is the anomalous magnetic moment of the free
electron, which is connected to the free-electron $g$-factor by $g_e \equiv 2(1 + \kappa) = 2 +
\alpha/\pi + \ldots$.

Many years ago Hegstrom \cite{hegstrom:73} derived the Hamiltonian describing the interaction of an
atom with the magnetic field, which accounts for the leading relativistic, QED, and nuclear recoil
effects. The resulting Hamiltonian is complete through orders of $\alpha^2$, $\alpha^3$,
$\alpha^2\,m/M$, and $\alpha^3\,m/M$. The corresponding numerical calculations for Li-like atoms
were performed by Yan~\cite{yan:01:prl,yan:02:jpb}.

In the present work, we address the leading relativistic and QED corrections to the $g$-factor of
of Li-like atoms. These corrections are induced by the effective Hamiltonian $\delta H$, which, for
the case of the $S$ states, can be simplified to take a very compact form,
\begin{eqnarray}
\delta H &=& \sum_a \mu_B\,Q_a\,\vec\sigma_a\cdot\vec B\,,\\
Q_a &=& Q_a^{(2)} + \kappa\,Q_a^{(3)}\,, \label{eqaa1} \\
Q_a^{(2)} &=& \frac{1}{3}\,\biggl(-2\,p_a^2+\frac{Z}{r_a}-\sum_{b\neq a}\frac{1}{r_{ab}}\biggr)\,, \\
Q_a^{(3)} &=& \frac{1}{3}\,\biggl(-\frac{p_a^2}{2}+\frac{Z}{r_a}-\sum_{b\neq a}\frac{1}{r_{ab}}\biggr)\,,
\end{eqnarray}
where the indices $a$ and $b = (1,2,3)$ numerate the electrons in the atom.

\begin{widetext}

Expectation values of the operators $Q_a$ are evaluated with the nonrelativistic wave function
$\psi$. This function is the antisymmetrized product $({\cal A})$ of the spacial function $\phi$
and the spin function $\chi$,
\begin{eqnarray}
\psi &=& {\cal A}[\phi(\vec r_1,\vec r_2,\vec r_3)\,\chi]\,,\\
\chi &=& [\alpha(1)\,\beta(2)-\beta(1)\,\alpha(2)]\,\alpha(3)\,,
\end{eqnarray}
where $\sigma_z\,\alpha(.) = \alpha(.) $ and $\sigma_z\,\beta(.) = -\beta(.) $. Matrix elements of
a spin-independent operator $H$, after eliminating spin variables, can be expressed as

\begin{align}
\langle \psi'|H|\psi\rangle = \Bigl\langle \phi'(r_1,\,r_2,\,r_3)\bigl|\,H\,\bigl|
 2\,\phi(r_1,r_2,r_3)+2\,\phi(r_2,r_1,r_3)
-\phi(r_2,r_3,r_1)-\phi(r_3,r_2,r_1)-\phi(r_3,r_1,r_2)-\phi(r_1,r_3,r_2) \Bigr\rangle\,.
\end{align}
Matrix elements of the spin-dependent operators are expressed as
\begin{equation}
\langle\psi'|\sum_a\,Q_a\,\vec \sigma_a|\psi\rangle
= \sum_a\langle\phi'| Q_a|\phi\rangle_F\,2\,\vec S\,,
\end{equation}
where $\vec S = \sum_a \vec{s}_a$ and
\begin{align}
\sum_a\langle \phi'|Q_a|\phi\rangle_F =&\
\Bigl\langle \phi'(r_1,\,r_2,\,r_3)\Bigl|2\,Q_3\,[\phi(r_1,r_2,r_3)+\phi(r_2,r_1,r_3)]
-(Q_1-Q_2+Q_3)\,[\phi(r_2,r_3,r_1)+\phi(r_3,r_2,r_1)]
\nonumber \\ &
-(Q_2-Q_1+Q_3)\,[\phi(r_1,r_3,r_2)+\phi(r_3,r_2,r_1)]\Bigr\rangle\,.
\end{align}
\end{widetext}

The corresponding corrections to the $g$-factor are
\begin{eqnarray} \label{eqa1}
g^{(2)}_{\infty} &=& 2\,\sum_a \bigl<  Q_a^{(2)}\bigr>_F\,,\\
g^{(3)}_{\infty} &=& \frac1{\pi} \sum_a \bigl<  Q_a^{(3)}\bigr>_F\,.
\label{eqa1a}
\end{eqnarray}

We calculated the matrix elements (\ref{eqa1}) and (\ref{eqa1a}) by using accurate variational wave
functions in the Hyleraas basis. The method is described in our previous investigations
\cite{puchalski:06,puchalski:08,puchalski:15}. Our numerical results for $g^{(2)}_{\infty}$ and
$g^{(3)}_{\infty}$ are presented in Tables~\ref{tab:g2} and \ref{tab:g3}, respectively. The values
listed in the tables were obtained by using the basis with the expansion parameter $\Omega =
n_1+n_2+n_3+n_4+n_5+n_6 = 12$. The specified uncertainties were obtained by taking the differences
of the results with $\Omega = 12$ and $11$. For lithium, we find good agreement with the previous
calculations by Yan~\cite{yan:01:prl,yan:02:jpb}, our results being several digits more accurate.
For lithium-like ions we observe small deviations outside of the Yan's error bars.

\subsection{Relativistic correction $\bm{g^{(2)}}_{\infty}$}
\label{sec:g2}

The leading relativistic correction of order $\alpha^2$ can be expanded in $1/Z$ as follows
\begin{align} \label{eq21}
g^{(2)}_{\infty}(Z) =  -\frac{Z^2}{6} + \frac{940}{2187}\,Z + c^{(2,0)}  + \frac{H^{(2,-1)}(Z)}{Z}\,,
\end{align}
where $H^{(2,-1)}(Z)$ is the remainder function that incorporates all higher orders in $1/Z$,
$H^{(2,-1)}(Z) \to c^{(2,-1)}$ as $Z\to \infty$. The leading coefficient of the expansion
(\ref{eq21}) follows from the hydrogenic limit summarized in Appendx~\ref{app:hydr}, whereas the
second coefficient is derived in the present work. The higher-order coefficients were obtained by
fitting the numerical results for $g^{(2)}_{\infty}$, as described in Appendix~\ref{app:fitting}.

Our fitting results for the first higher-order expansion coefficients are
\begin{align} \label{eq23}
c^{(2,0)}  = -0.128\,204\,(9)\,, \ \ \
c^{(2,-1)}  = 0.028\,78\,(46)\,.
\end{align}
We would like to stress that in order to achieve such precision of the fitted coefficients, it was
important to have highly accurate numerical results for $g^{(2)}_{\infty}$ for a sufficiently wide
range of $Z$. In particular, if we apply the same fitting procedure to the analogous results of
Yan~\cite{yan:01:prl,yan:02:jpb}, we get results consistent with Eq.~(\ref{eq23}) but much less
accurate.

Using the result for $c^{(2,0)}$ from Eq.~(\ref{eq23}), we can extract the remainder function
$H^{(2,-1)}(Z)$ from our numerical data for $g^{(2)}_{\infty}$. The corresponding results are
presented in the last column of Table~\ref{tab:g2}. The errors of the listed values of $H^{(2,-1)}$
come from the uncertainty of $c^{(2,0)}$. In the case of silicon, we obtain $H^{(2,-1)}(14) =
0.024\,77\,(13)\,,$ which agrees with but is more precise than the corresponding result of
$0.024\,4\,(15)$ obtained by the Configuration-Interaction Dirac-Fock (CI-DF) method in
Ref.~\cite{volotka:14}. We note that $H^{(2,-1)}$ previously yielded one of the two main errors of
the total theoretical $g$-factor predictions.

\subsection{QED correction $\bm{g^{(3)}}_{\infty}$}
\label{sec:g3}

The leading QED correction  of order $\alpha^3$ can be expanded in $1/Z$ as follows
\begin{align} \label{eq25}
g^{(3)}_{\infty}(Z) =  \frac{1}{24 \pi}\,Z^2 -\frac{274}{2187 \pi}\,Z + H^{(3,0)}(Z)\,,
\end{align}
where $H^{(3,0)}(Z)$ is the remainder that incorporates the higher orders in $1/Z$, $H^{(3,0)} \to
c^{(3,0)}$ as $Z\to \infty$. The leading coefficient of the expansion (\ref{eq25}) comes from the
hydrogenic limit, Eq.~(\ref{eq01}), whereas the second term was derived in
Ref.~\cite{glazov:04:pra}.

Using the known results for the first two terms of the expansion (\ref{eq25}), we identify values
of the remainder function $H^{(3,0)}(Z)$ from our numerical results for $g^{(3)}_{\infty}$, with
the corresponding results presented in the last column of Table~\ref{tab:g3}. In particular, for
silicon we obtain $H^{(3,0)}(14) = 0.022\,467\,9\,,$ which agrees with the corresponding value of
$0.0224\,(10)$, obtained in Ref.~\cite{glazov:04:pra} by fitting the results of
Yan~\cite{yan:01:prl,yan:02:jpb}.

Our fitting results for the first expansion coefficients of $H^{(3,0)}(Z)$ are
\begin{align} \label{eq28}
c^{(3,0)} = 0.022\,412\,(2)\,, \ \
c^{(3,-1)} = 0.000\,53\,(7)\,.
\end{align}
These results can be used for estimating the $H^{(3,0)}(Z)$ function for higher values of $Z$.

We note that $g^{(3)}_{\infty}$ is induced by the one-loop part of the anomalous magnetic moment
(AMM), $\alpha/\pi$. According to Eq.~(\ref{eqaa1}), analogous corrections due to the $n$-loop part
of the AMM differ from $g^{(3)}_{\infty}$ only by a prefactor. In particular, the two-loop part of
$g^{(4)}_{\infty}$ is
\begin{align}
g^{(4)}_{\rm twoloop} = \frac{2\,A_2}{\pi}\,g^{(3)}_{\infty}\,,
\end{align}
where $A_2$ is the two-loop contribution to the AMM defined in Eq.~(\ref{eq02}).

\subsection{Recoil correction $\bm{g^{(2)}_{M}}$}
\label{sec:g2M}

The leading recoil correction  of order $\alpha^2\,m/M$ can be expanded in $1/Z$ as follows
\begin{align} \label{eq29}
g^{(2)}_{M}(Z) =  \frac{1}{4}\,Z^2 + Z\,H^{(2,1)}_M(Z)\,,
\end{align}
where the leading coefficient follows from the hydrogenic limit, Eq.~(\ref{eq04}), and
$H^{(2,1)}_M(Z)$ is the higher-order remainder function, $H^{(2,1)}_M \to c^{(2,1)}_M$ as $Z\to
\infty$.

In the present work we obtain the remainder function and the coefficient $c^{(2,1)}_M$ by fitting
the results of Yan~\cite{yan:01:prl,yan:02:jpb}. Our value for the coefficient
\begin{align} \label{eq31}
c^{(2,1)}_M &\ = -0.860\,3\,(8)\,
\end{align}
disagrees with the corresponding result of $-0.825\,(5)$ from Ref.~\cite{glazov:04:pra} obtained by
fitting the same results of Yan. We do not know the reason for this disagreement. Our fitting
procedure was the same as used for the $g^{(2)}_{\infty}$ and $g^{(3)}_{\infty}$ corrections and it
reproduces well the analytical value of the leading coefficient in Eq.~(\ref{eq29}). We also obtain
the remainder function for silicon as
\begin{align} \label{eq31}
H^{(2,1)}_M(14) &\ = -0.832\,9\,(1)\,.
\end{align}

\subsection{Radiative recoil correction $\bm{g^{(3)}_{M}}$}
\label{sec:g3M}

The radiative recoil correction of order $\alpha^3\,m/M$ can be expanded in $1/Z$ as follows
\begin{align} \label{eq40}
g^{(3)}_{M}(Z) =  -\frac{1}{12\pi}\,Z^2 + Z\,H^{(3,1)}_M(Z)\,,
\end{align}
where the leading coefficient follows from the hydrogenic limit, Eq.~(\ref{eq04}), and
$H^{(3,1)}_M(Z)$ is the higher-order remainder, $H^{(3,1)}_M \to c^{(3,1)}_M$ as $Z\to \infty$.

In the present work we obtain the remainder function and the coefficient $c^{(3,1)}_M$ by fitting
the results of Yan~\cite{yan:01:prl,yan:02:jpb}. Our values for the coefficient and the remainder
are
\begin{align} \label{eq41}
c^{(3,1)}_M  =  0.040\,23\,(4)\,,\ \
H^{(3,1)}_M(14)  = 0.040\,337\,(6)\,.
\end{align}

\begin{table}
\caption{The leading relativistic contribution $g^{(2)}_{\infty}$ to the $g$-factor of the ground state of Li-like atoms
and the corresponding higher-order remainder function $H^{(2,-1)}$ defined by Eq.~(\ref{eq21}).
\label{tab:g2}}
\begin{ruledtabular}
\begin{tabular}{ldd}
% $Z$ &    \multicolumn{1}{c}{g^{(2)}_{\infty}}   &   \multicolumn{1}{c}{$H^{(2,-1)}$ }\\
 $Z$ & \multicolumn{1}{c}{$g^{(2)}_{\infty}$} & \multicolumn{1}{c}{$H^{(2,-1)}$ }\\\hline\\[-5pt]
  3 & -0.343\,332\,404\,(3) & -0.013\,70\,(3) \\
    & -0.343\,332\,42\,(7)\,^a\\
  4 & -1.074\,312\,532\,(7) & 0.005\,23\,(4) \\
  5 & -2.143\,265\,913\,(2) & 0.012\,71\,(5) \\
  6 & -3.546\,553\,940\,(2) & 0.016\,65\,(5) \\
  7 & -5.283\,459\,746\,(4) & 0.019\,06\,(6) \\
  8 & -7.353\,784\,626\,(1) & 0.020\,69\,(7) \\
  9 & -9.757\,463\,309\,(3) & 0.021\,85\,(8) \\
 10 & -12.494\,472\,721\,(5) & 0.022\,73\,(9) \\
 11 & -15.564\,804\,889\,(4) & 0.023\,4\,(1) \\
 12 & -18.968\,457\,638\,(1) & 0.024\,0\,(1) \\
    & -18.968\,460\,5\,(2)\,^a\\
 13 & -22.705\,431\,064\,(2) & 0.024\,4\,(1) \\
 14 & -26.775\,726\,109\,(1) & 0.024\,8\,(1) \\
\end{tabular}
\end{ruledtabular}
$^a$ Ref.~\cite{yan:02:jpb}.
\end{table}

\begin{table}
\caption{The leading QED contribution $g^{(3)}_{\infty}$ to the $g$-factor of the ground state of Li-like atoms
and the corresponding higher-order remainder function $H^{(3,0)}$ defined by Eq.~(\ref{eq25}).
\label{tab:g3}}
\begin{ruledtabular}
\begin{tabular}{ldd}
 $Z$ & \multicolumn{1}{c}{$g^{(3)}_{\infty}$} & \multicolumn{1}{c}{$H^{(3,0)}$ }\\\hline\\[-5pt]
  3 & 0.023\,071\,092\,3\,(7) & 0.023\,344 \\
  &   0.023\,071\,11\,(2)\,^a\\
  4 & 0.075\,560\,527\,2\,(1) & 0.022\,873 \\
  5 & 0.154\,876\,875\,2\,(2) & 0.022\,703 \\
  6 & 0.260\,805\,551\,9\,(3) & 0.022\,619 \\
  7 & 0.393\,295\,230\,1\,(5) & 0.022\,570 \\
  8 & 0.552\,328\,059\,8\,(1) & 0.022\,539 \\
  9 & 0.737\,896\,374\,9\,(2) & 0.022\,518 \\
 10 & 0.949\,996\,389\,1\,(4) & 0.022\,502 \\
 11 & 1.188\,626\,037\,9\,(1) & 0.022\,490 \\
 12 & 1.453\,784\,110\,3\,(1) & 0.022\,481 \\
    & 1.453\,784\,66\,(4)\,^a\\
 13 & 1.745\,469\,851\,3\,(1) & 0.022\,474 \\
 14 & 2.063\,682\,768\,1\,(1) & 0.022\,468 \\
\end{tabular}
\end{ruledtabular}
$^a$ Ref.~\cite{yan:02:jpb}.
\end{table}

\section{Results and discussion}

The summary of individual binding corrections to the $g$-factors of Li-like silicon, oxygen, and
carbon ions is presented in  Tables~\ref{tab:si}, \ref{tab:o} and \ref{tab:c}, respectively. The
sum of all binding corrections gives the difference between the $g$-factor of the atom and the
free-electron $g$-factor, $g - g_e$, which may be compared to the experimental data and to other
theoretical predictions by using the experimental value of the free-electron $g$-factor
\cite{hanneke:08},
\begin{align}
g_e = 2.002\,319\,304\,361\,(6)\,.
\end{align}

In the tables, the columns labelled ``LO'' present results for the lowest-order (in $\Za$) parts of
the corresponding corrections. The columns labelled ``HO'' contain results for the higher-order
remainders, which are suppressed by a factor of $(\Za)^2$ as compared to the corresponding LO part.

The largest contribution to $g-g_e$ comes from the electron-electron interaction. The corresponding
LO part is discussed in Sec.~\ref{sec:g2}. The $1/Z^0$ HO term comes from the hydrogenic limit,
Eq.~(\ref{eq00}). The $1/Z^1$ HO term originates from the one-photon exchange diagrams, first
calculated in Ref.~\cite{shabaev:02:li} and reevaluated in this work to a higher precision. The
$1/Z^2$ HO term comes from the two-photon exchange diagrams, which were calculated to all orders in
$\Za$ in Ref.~\cite{volotka:14}. For silicon, we identify the $1/Z^2$ HO term from the all-order
numerical result of Ref.~\cite{volotka:14}. For oxygen and carbon, there were no results reported
in there, so we estimate the $1/Z^2$ HO term by scaling the silicon's result. E.g., for oxygen we
obtain
\begin{align} \label{eq001}
\delta g = -0.000\,92\,\alpha^2\,(8/14)^2 = -0.000\,3\,\alpha^2\,.
\end{align}
We ascribe the uncertainty of 50\% to this estimation. The $1/Z^3$ HO term is unknown; the
corresponding uncertainty was estimated as the $1/Z^3$ LO term multiplied by the ratio of the
$1/Z^2$ HO-to-LO terms, and by a conservative factor of 1.5.

The LO part of the one-loop QED correction is discussed in Sec.~\ref{sec:g3}. The corresponding
$1/Z^0$ HO term comes from the hydrogenic limit, Eq.~(\ref{eq01}). The $1/Z^1$ HO term is induced
by the screened QED diagrams, calculated to all orders in $\Za$ in
Refs.~\cite{glazov:10,volotka:14}. For silicon, we take the result presented in Table II of
Refs.~\cite{volotka:14} and identify the $1/Z^1$ contribution by subtracting the $1/Z^{2+}$ part
taken from Table~V of Ref.~\cite{glazov:04:pra}. For carbon and oxygen, we scale the silicon result
and ascribe a 100\% uncertainty to this estimate. The $1/Z^2$ HO term has not been evaluated yet.
We estimated its uncertainty as the $1/Z^2$ LO term multiplied by the ratio of the $1/Z^1$ HO and
LO terms, and by an additional conservative factor of 1.5.

The LO part of the recoil correction is discussed in Sec.~\ref{sec:g2M}. The only HO recoil
contribution available today for oxygen and silicon is the $(\Za)^4m/M$ correction obtained in
Ref.~\cite{glazov:10} in  the hydrogenic limit (see Eq.~(\ref{eq04})). We note that a calculation
complete to all orders in $\Za$ was reported in Ref.~\cite{sturm:16}, but only for calcium. In the
absence of such calculations for other ions, we estimate the uncertainty due to higher orders in
$\Za$ on the basis of the results available for the $1s$ state \cite{shabaev:02:recprl}.

The $1/Z^0$ part of the two-loop QED correction is given by Eq.~(\ref{eq02}), whereas the
$1/Z^{1+}$ part is described in Sec.~\ref{sec:g3}. The finite nuclear size correction is taken from
our previous investigation \cite{yerokhin:16:gfact:pra}.

In Tables~\ref{tab:si}, \ref{tab:o} and \ref{tab:c}, we summarize all known theoretical
contributions to $g-g_e$ for Li-like silicon, oxygen, and carbon and compare the results with
previous theoretical and experimental data. For silicon, we observe a very good agreement with the
theoretical prediction by Volotka et al.~\cite{volotka:14} and with the experimental result
\cite{wagner:13}. Our prediction is slightly more accurate than that by Volotka et al., mainly
because of the improvement in the $1/Z^{3+}$ electron-correlation correction. For oxygen and
carbon, we find a marginal agreement with the previous theoretical calculations of Glazov et
al.~\cite{glazov:04:pra} but improve their accuracy by a factor of 3 (oxygen) or 4 (carbon). The
main difference between the results comes from the $1/Z^{2+}$ electron-correlation correction,
which was evaluated by the CI-DF method in Ref.~\cite{glazov:04:pra} and by the NRQED method in the
present work.

The largest uncertainty of our theoretical prediction for silicon stems from the $1/Z^1$ part of
the one-loop QED effect, also known as the screened QED correction. The corresponding uncertainty
is the estimated error of the numerical evaluation \cite{volotka:14}, which can be improved by
dedicated calculations. For oxygen and carbon, the largest theoretical error comes from the $1/Z^2$
part of the electron-electron interaction correction. This error can be eliminated by extending the
all-order calculation of the two-photon exchange diagrams by Volotka et al.~\cite{volotka:14} to
lower-$Z$ ions, or by performing the NRQED calculations of the next-order $\alpha^4$ effect.

Summing up, we have performed NRQED calculations of the electron-correlation effects to the $g$
factor of the ground state of Li-like atoms. By fitting the $Z$ dependence of the NRQED results for
the $\alpha^2$ and $\alpha^3$ effects and the corresponding recoil corrections, we have identified
their individual $1/Z$-expansion contributions. Combining the obtained data with the results of the
all-order (in $\Za$) calculations performed within the $1/Z$ expansion, we have derived unified
theoretical predictions for the $g$-factor of light Li-like ions and improved the theoretical
precision.

\section*{Acknowledgement}

V.A.Y. acknowledges support by the Ministry of Education and Science of the Russian Federation
Grant No. 3.5397.2017/BY. Work of M.P. and K.P. was supported by the National Science Center
(Poland) Grant No. 2012/04/A/ST2/00105. Fruitful discussions with D.~A.~Glazov, V.~M.~Shabaev, and
A.~V.~Volotka are gratefully acknowledged.

\appendix

\section{Hydrogenic limit}
\label{app:hydr}

The $g$-factor of the ground state of a Li-like atom in the hydrogenic limit (i.e., neglecting the
electron-electron interaction) coincides with the $g$-factor of the $2s$ state of the corresponding
H-like ion. In this section we summarize the theory of the $g$-factor of the hydrogenic $2s$ state.

The relativistic value of the $2s$ $g$-factor  is obtained from the Dirac equation, with the
(point-nucleus) result
\begin{align}\label{eq00}
g = \frac23 \left[ 1 + \sqrt{2+2\sqrt{1-(\Za)^2}}\,\right] = 2 - \frac{(\Za)^2}{6}+ \ldots\,.
\end{align}

The one-loop QED correction (for the point nucleus) is
\cite{grotch:71,pachucki:04:prl,pachucki:05:gfact}
\begin{align}\label{eq01}
g_{\rm QED}^{(1)} = \frac{\alpha}{\pi}\biggl\{  &\  1 + \frac{(\Za)^2}{24} + \frac{(\Za)^4}{8}\biggl[
 \frac{32}{9}\,\ln[(\Za)^{-2}] + b_{40}^{(1)} \biggr]
 \nonumber \\ &
 + \frac{(\Za)^5}{8}\,H^{(1)}(\Za)\biggr\}\,,
\end{align}
where $b_{40}^{(1)} = -11.774\,382\,27$ \cite{pachucki:04:prl,pachucki:05:gfact} and $H(\Za)$ is
the remainder function that incorporates all higher orders in $\Za$. The self-energy part of the
remainder function was obtained numerically in Ref.~\cite{yerokhin:tobe},
\begin{align}
H^{(1)}_{\rm SE}(6\alpha) = 22.48\,(1)\,,\nonumber \\
H^{(1)}_{\rm SE}(8\alpha) = 22.221\,(4)\,,\nonumber \\
H^{(1)}_{\rm SE}(14\alpha) = 21.486\,(1)\,.
\end{align}
The vacuum-polarization part of the remainder function consists of the so-called electric-loop and
magnetic-loop parts. The electric-loop part is relatively simple and evaluated numerically by many
authors, e.g., by us,
\begin{align}
H^{(1)}_{\rm VP,el}(6\alpha) &\ = 1.46\,,\nonumber \\
H^{(1)}_{\rm VP,el}(8\alpha) &\ = 1.388\,,\nonumber \\
H^{(1)}_{\rm VP,el}(14\alpha) &\ = 1.199\,6\,,
\end{align}
whereas the magnetic-loop part is given by \cite{karshenboim:05,lee:05}
\begin{align}
H^{(1)}_{\rm VP,ml}(\Za) &\ = \frac{7\pi}{216} - (\Za) \frac{8}{135} \biggl[\ln (\Za) + 2.6 + \frac58\biggr]\,.
\end{align}

The two-loop QED correction is
\begin{align}\label{eq02}
g_{\rm QED}^{(2)} = \frac{\alpha^2}{\pi^2}\biggl\{ &\, 2A_2 + 2A_2\frac{(\Za)^2}{24} + \frac{(\Za)^4}{8}\biggl[
 \frac{28}{9}\,\ln[(\Za)^{-2}]
 \nonumber \\ &
 + b_{40}^{(2)} + \frac{16-19\pi^2}{108} \biggr]\biggr\}\,,
\end{align}
where $A_2 = -0.328\,478\,444\,00\ldots$ is the two-loop contribution to the electron anomalous
magnetic moment and $b_{40}^{(2)} = -17.157\,236\,58$~\cite{pachucki:04:prl,pachucki:05:gfact}. The
last term $O((\Za)^4)$ in Eq.~(\ref{eq02}) is the light-by-light scattering contribution recently
calculated in Ref.~\cite{czarnecki:16}.

The three and higher-loop QED corrections can be summarized as
\begin{align}\label{eq03}
g_{\rm QED}^{(\ge3)} = \sum_{n = 3}^{5} \frac{\alpha^n}{\pi^n}\biggl\{ &\, 2A_n + 2A_n\frac{(\Za)^2}{24}\biggr\}\,,
\end{align}
where \cite{aoyama:15,laporta:17}
\begin{align*}
A_3 &\ = 1.181\,234\,017\ldots\,,\\
A_4 &\ = -1.912\,245\,765\ldots\,,\\
A_5 &\ = 7.79\,(34)\,.
\end{align*}

The recoil correction, including the second-order recoil $O((m/M)^2)$ and the radiative recoil
$O(\alpha\, m/M)$, is \cite{grotch:71,faustov:70:plb,close:71,glazov:04:pra}
\begin{align}\label{eq04}
g_{\rm rec} = \frac{m}{M}\frac{(\Za)^2}{4}\biggl[ 1 + \frac{11}{48} (\Za)^2
 - \frac{m}{M} (1+Z) - \frac{\alpha}{3\pi}\biggr]\,.
\end{align}

The finite nuclear size correction including the corresponding QED contribution is
\begin{align}
g_{\rm N} & = \frac{2}{5}\,\bigl(\Za R_{\rm sph})^{2\gamma}\frac{(\Za)^2}{2}\biggl[1 + (\Za)^2 H_{\rm N}^{(0,2+)}\biggr]
\nonumber \\ & \times
\biggl[1 + \frac{\alpha}{\pi}\, G_{\rm NQED}\biggr]\,,
\end{align}
where $\gamma = \sqrt{1-(\Za)^2}$ and $R_{\rm sph} = \sqrt{5/3}\,R$ the radius of the nuclear
sphere with the root-mean-square radius $R$, and the remainder functions $H_{\rm N}^{(0,2+)}$ and
$G_{\rm NQED}$ were evaluated in
Refs.~\cite{yerokhin:13:jpb,yerokhin:16:gfact:prl,yerokhin:16:gfact:pra}.

\section{Fitting of the $\bm{1/Z}$ expansion coefficients}
\label{app:fitting}

In this section we describe the fitting procedure used for the identification of the coefficients
of the $1/Z$ expansion.

The general task is to fit a data set of $n$ points $(Z_i,F_i)$, $i = 1,\ldots,n$, to the following
model function with $N$ ($N<n$) parameters,
\begin{align}
f_N(Z) = \sum_{k = 0}^{N-1} c_k\, Z^{a-k}\,,
\end{align}
where $a$ is the exponent of the leading term of the $1/Z$ expansion.

In order to find the optimal values of the fitting parameters $c_k$, we use the weighted
least-squares regression. Specifically, we minimize the functional
\begin{align}
S_N = \sum_{i = 1}^{n} \frac{\bigl(F_i - f_N(Z_i)\bigr)^2}{\delta F_i^2 + \sigma_N^2(Z_i)}\,,
\end{align}
where $\delta F_i$ are the numerical errors of $F_i$ and $\sigma_N(Z)$ is the estimate of the error
due to the truncation of the $1/Z$ expansion in the fitting function, taken as the last term of the
fitting ansatz divided by $Z$,
\begin{align}
\sigma_N(Z) = c_{N-1}\,Z^{a-N}\,.
\end{align}
In practice, we make our fit in two steps. First, we perform the least-square regression without
weights. The obtained value of the $c_{N-1}$ coefficient is then used for the estimation of the
truncation error in the weighted least-square regression performed on the second step.

In the cases relevant for the present work, one or two first coefficients of the $1/Z$ expansion
are known analytically. We use this fact in order to access the errors of our fitting procedure.
First, we treat one of the known coefficients as a free fitting parameter and select three
different fitting functions that give the best approximation to the known result. After that, we
set the known coefficients to their exact values, perform the fit with the three fitting functions,
and finally take the average of the three results and the maximal deviation between them as the
final value and its error, respectively.

Fitting the results for the $\alpha^2$ and $\alpha^3$ corrections (whose numerical accuracy is very
high), we used fitting functions with 8-10 parameters. For the recoil corrections (whose accuracy
is much lower), we used 5-6 fitting parameters.

%
%============================================
% Li-like Silicon:
% Z =  14
% A =  28
%  M/m =     50984.83273
%  RMS =   3.1224 (0.0024)  fm
%
\begin{table*}
\caption{Binding corrections to the $g$-factor of the ground state of $^{28}$Si$^{ 11+}$.
``LO'' denotes the lowest-order (in $\Za$) contribution of the corresponding correction, to be multiplied by the prefactor specified in the second column.
``HO'' denotes the higher-order (in $\Za$) contribution, to be multiplied by the prefactor specified in the second column.
$\delta g$ denotes the contribution to the $g$-factor, obtained as the sum of the lowest-order and the higher-order contributions, multiplied by the prefactor.
Nuclear parameters used in the calculation are: $M/m =     50\,984.832\,73$ and $R =
3.1224\,(24)$~fm.
\label{tab:si}}
\begin{ruledtabular}
\begin{tabular}{ccddd}
 \multicolumn{1}{c}{Order}   & \multicolumn{1}{c}{Prefactor}  & \multicolumn{1}{c}{LO}  & \multicolumn{1}{c}{HO}  & \multicolumn{1}{c}{$\delta g \times 10^6$} \\
\hline\\[-5pt]
\multicolumn{1}{l}{Electron-electron interaction:} \\
$1/Z^0$      & $\alpha^2 Z^2$                           & -0.166\,666\,667 & -0.000\,546\,606 & -1\,745.249\,323 \\
$1/Z^1$      & $\alpha^2 Z$                             & 0.429\,812\,529 & 0.001\,552\,492 & 321.590\,803 \\
$1/Z^2$      & $\alpha^2$                               & -0.128\,204\,(9) & -0.000\,92\,(1) & -6.876\,0\,(5) \\
$1/Z^{3+}$   & $\alpha^2\,Z^{-1}$                       & 0.024\,8\,(1) & 0.000\,0\,(3) & 0.094\,(1) \\
\multicolumn{1}{l}{One-loop QED:} \\
$1/Z^0$      & $\alpha^3\,Z^2$                          & 0.013\,262\,912 & 0.002\,813\,49\,(4) & 1.224\,449\,(3) \\
$1/Z^1$      & $\alpha^3\,Z$                            & -0.039\,879\,702 & -0.005\,1\,(9) & -0.245\,(5) \\
$1/Z^{2+}$   & $\alpha^3$                               & 0.022\,467\,862 & 0.000\,(4) & 0.009\,(2) \\
\multicolumn{1}{l}{Recoil:} \\
$1/Z^0$      & $\alpha^2\,(m/M)\,Z^2$                   & 0.250\,000\,000 & 0.001\,(8) & 0.051\,(2) \\
$1/Z^{1+}$   & $\alpha^2\,(m/M)\,Z$                     & -0.832\,9\,(1) & 0.00\,(4) & -0.012\,2\,(6) \\
\multicolumn{1}{l}{Two-loop QED:} \\
$1/Z^0$      & $\alpha^4\,Z^2$                          & -0.002\,773\,485 & -0.000\,6\,(6) & -0.001\,9\,(3) \\
$1/Z^1$      & $\alpha^4\,Z$                            & 0.008\,339\,479 & 0.000\,(3) & 0.000\,3\,(1) \\
\multicolumn{1}{l}{Finite nuclear size:} \\
$1/Z^0$      & $(\Za R_{\rm sph})^{2\gamma}\alpha^2Z^2$ & 0.200\,000\,000 & 0.001\,90\,(7) & 0.002\,574\,(4) \\
$1/Z^1$      & $(\Za R_{\rm sph})^{2\gamma}\alpha^2 Z$  & -0.570\,2\,(2) & -0.008\,0\,(3) & -0.000\,527\,(1) \\
$1/Z^{2}$    & $(\Za R_{\rm sph})^{2\gamma}\alpha^2 $   & 0.214\,(5) & 0.002\,(3) & 0.000\,014 \\
\multicolumn{1}{l}{Radiative recoil:} \\
$1/Z^0$      & $\alpha^3\,(m/M)\,Z^2$                   & -0.026\,525\,824 & 0.000\,(1) & -0.000\,040\,(2) \\
$1/Z^{1+}$   & $\alpha^3\,(m/M)\,Z$                     & 0.040\,337\,(6) & 0.000\,(2) & 0.000\,004 \\
\multicolumn{1}{l}{Second-order recoil:} \\
$1/Z^0$      & $\alpha^2\,(m/M)^2\,Z^2$                 & -3.750\,000\,000 & 0.0\,(2) & -0.000\,015\,(1) \\
\multicolumn{1}{l}{$\ge$3-loop QED:} \\
$1/Z^0$      & $\alpha^5\,Z^2$                          & 0.0031\,629\,03 & 0.000\,0\,(7) & 0.000\,013\,(3) \\
\multicolumn{1}{l}{Total $g-g_e$:} \\
\multicolumn{1}{l}{Theory, this work}     &                                          &  &  &  -1\,429.412\,(6) \\
%$g$          &                                          &  &  & 2000889.892\,(6) \\
\multicolumn{1}{l}{Theory \cite{volotka:14} }  &                              &  &  & -1\,429.412\,(8) \\
\multicolumn{1}{l}{Experiment \cite{wagner:13} } &                            &  &  & -1\,429.414\,5\,(21)
\end{tabular}
\end{ruledtabular}
\end{table*}

\begin{table*}
\caption{Individual binding correction to the $g$-factor of the ground state of $^{16}$O$^{5+}$.
Notations are the same as in Table~\ref{tab:si}.
Nuclear parameters used in the calculation are: $M/m =     29\,148.949\,75$ and $R =   2.6991\,(52)$~fm.
\label{tab:o}}
\begin{ruledtabular}
\begin{tabular}{ccddd}
 \multicolumn{1}{c}{Order}   & \multicolumn{1}{c}{Prefactor}  & \multicolumn{1}{c}{LO}  & \multicolumn{1}{c}{HO}  & \multicolumn{1}{c}{$\delta g \times 10^6$} \\
\hline\\[-5pt]
\multicolumn{1}{l}{Electron-electron interaction:} \\
$1/Z^0$      & $\alpha^2 Z^2$                           & -0.166\,666\,667 & -0.000\,177\,823 & -568.620\,484 \\
$1/Z^1$      & $\alpha^2 Z$                             & 0.429\,812\,529 & 0.000\,505\,635 & 183.320\,201 \\
$1/Z^2$      & $\alpha^2$                               & -0.128\,204\,(9) & -0.000\,3\,(2) & -6.843\,(8) \\
$1/Z^{3+}$   & $\alpha^2\,Z^{-1}$                       & 0.020\,69\,(7) & 0.000\,00\,(7) & 0.137\,7\,(7) \\
\multicolumn{1}{l}{One-loop QED:} \\
$1/Z^0$      & $\alpha^3\,Z^2$                          & 0.013\,262\,912 & 0.001\,330\,41\,(3) & 0.362\,936\,(1) \\
$1/Z^1$      & $\alpha^3\,Z$                            & -0.039\,879\,702 & -0.002\,(2) & -0.129\,(5) \\
$1/Z^{2+}$   & $\alpha^3$                               & 0.022\,539\,315 & 0.000\,(1) & 0.008\,8\,(5) \\
\multicolumn{1}{l}{Recoil:} \\
$1/Z^0$      & $\alpha^2\,(m/M)\,Z^2$                   & 0.250\,000\,000 & 0.000\,(2) & 0.029\,3\,(2) \\
$1/Z^{1+}$   & $\alpha^2\,(m/M)\,Z$                     & -0.811\,399\,(3) & 0.000\,(9) & -0.011\,9\,(1) \\
\multicolumn{1}{l}{Two-loop QED:} \\
$1/Z^0$      & $\alpha^4\,Z^2$                          & -0.002\,773\,485 & -0.000\,05\,(6) & -0.000\,51\,(1) \\
$1/Z^1$      & $\alpha^4\,Z$                            & 0.008\,339\,479 & 0.000\,0\,(2) & 0.000\,189\,(5) \\
\multicolumn{1}{l}{Finite nuclear size:} \\
$1/Z^0$      & $(\Za R_{\rm sph})^{2\gamma}\alpha^2Z^2$ & 0.200\,000\,000 & 0.000\,42\,(6) & 0.000\,194\,(1) \\
$1/Z^1$      & $(\Za R_{\rm sph})^{2\gamma}\alpha^2 Z$  & -0.570\,2\,(2) & -0.002\,4\,(3) & -0.000\,069 \\
$1/Z^{2}$    & $(\Za R_{\rm sph})^{2\gamma}\alpha^2 $   & 0.214\,(5) & 0.001\,(1) & 0.000\,003 \\
\multicolumn{1}{l}{Radiative recoil:} \\
$1/Z^0$      & $\alpha^3\,(m/M)\,Z^2$                   & -0.026\,525\,824 & 0.000\,0\,(3) & -0.000\,023 \\
$1/Z^{1+}$   & $\alpha^3\,(m/M)\,Z$                     & 0.040\,504\,18\,(6) & 0.000\,0\,(5) & 0.000\,004 \\
\multicolumn{1}{l}{Second-order recoil:} \\
$1/Z^0$      & $\alpha^2\,(m/M)^2\,Z^2$                 & -2.250\,000\,000 & 0.00\,(3) & -0.000\,009 \\
\multicolumn{1}{l}{$\ge$3-loop QED:} \\
$1/Z^0$      & $\alpha^5\,Z^2$                          & 0.003\,162\,903 & 0.000\,00\,(4) & 0.000\,004 \\
\multicolumn{1}{l}{Total $g-g_e$:} \\
\multicolumn{1}{l}{Theory, this work}     &                                          &  &  & -391.745\,9\,(96) \\
%$g$          &                                          &  &  & 2001927.5585\,(96) \\
\multicolumn{1}{l}{Theory \cite{glazov:04:pra}}  &                            &  &         & -391.700\,(32)
\end{tabular}
\end{ruledtabular}
\end{table*}

%
%============================================
% Li-like Carbon:
% Z =   6
% A =  12
%  M/m =     21868.66386
%  RMS =   2.4702 (0.0022)  fm
%
\begin{table*}
\caption{Individual binding correction to the $g$-factor of the ground state of $^{12}$C$^{3+}$.
Notations are the same as in Table~\ref{tab:si}.
Nuclear parameters used in the calculation are: $M/m =     21\,868.663\,86$ and $R =   2.4702\,(24)$~fm.
\label{tab:c}}
\begin{ruledtabular}
\begin{tabular}{ccddd}
 \multicolumn{1}{c}{Order}   & \multicolumn{1}{c}{Prefactor}  & \multicolumn{1}{c}{LO}  & \multicolumn{1}{c}{HO}  & \multicolumn{1}{c}{$\delta g \times 10^6$} \\
\hline\\[-5pt]
\multicolumn{1}{l}{Electron-electron interaction:} \\
$1/Z^0$      & $\alpha^2 Z^2$                           & -0.166\,666\,667 & -0.000\,099\,947 & -319.699\,730 \\
$1/Z^1$      & $\alpha^2 Z$                             & 0.429\,812\,529 & 0.000\,284\,265 & 137.419\,421 \\
$1/Z^2$      & $\alpha^2$                               & -0.128\,204\,(9) & -0.000\,17\,(8) & -6.836\,(4) \\
$1/Z^{3+}$   & $\alpha^2\,Z^{-1}$                       & 0.016\,65\,(5) & 0.000\,00\,(3) & 0.147\,8\,(6) \\
\multicolumn{1}{l}{One-loop QED:} \\
$1/Z^0$      & $\alpha^3\,Z^2$                          & 0.013\,262\,912 & 0.000\,879\,11\,(3) & 0.197\,838 \\
$1/Z^1$      & $\alpha^3\,Z$                            & -0.039\,879\,702 & -0.000\,9\,(9) & -0.095\,(2) \\
$1/Z^{2+}$   & $\alpha^3$                               & 0.022\,618\,936 & 0.000\,0\,(8) & 0.008\,8\,(3) \\
\multicolumn{1}{l}{Recoil:} \\
$1/Z^0$      & $\alpha^2\,(m/M)\,Z^2$                   & 0.250\,000\,000 & 0.000\,1\,(9) & 0.021\,93\,(8) \\
$1/Z^{1+}$   & $\alpha^2\,(m/M)\,Z$                     & -0.793\,800\,(1) & 0.000\,(4) & -0.011\,60\,(6) \\
\multicolumn{1}{l}{Two-loop QED:} \\
$1/Z^0$      & $\alpha^4\,Z^2$                          & -0.002\,773\,485 & 0.000\,02\,(2) & -0.000\,281(2) \\
$1/Z^1$      & $\alpha^4\,Z$                            & 0.008\,339\,479 & 0.000\,00\,(8) & 0.000\,142\,(1) \\
\multicolumn{1}{l}{Finite nuclear size:} \\
$1/Z^0$      & $(\Za R_{\rm sph})^{2\gamma}\alpha^2Z^2$ & 0.200\,000\,000 & 0.000\,18\,(9) & 0.000\,051 \\
$1/Z^1$      & $(\Za R_{\rm sph})^{2\gamma}\alpha^2 Z$  & -0.570\,2\,(2) & -0.001\,4\,(7) & -0.000\,024 \\
$1/Z^{2}$    & $(\Za R_{\rm sph})^{2\gamma}\alpha^2 $   & 0.214\,(5) & 0.000\,4\,(8) & 0.000\,002 \\
\multicolumn{1}{l}{Radiative recoil:} \\
$1/Z^0$      & $\alpha^3\,(m/M)\,Z^2$                   & -0.026\,525\,824 & 0.000\,0\,(1) & -0.000\,017 \\
$1/Z^{1+}$   & $\alpha^3\,(m/M)\,Z$                     & 0.040\,698\,000 & 0.000\,0\,(2) & 0.000\,004 \\
\multicolumn{1}{l}{Second-order recoil:} \\
$1/Z^0$      & $\alpha^2\,(m/M)^2\,Z^2$                 & -1.750000000 & 0.00\,(1) & -0.000007 \\
\multicolumn{1}{l}{$\ge$3-loop QED:} \\
$1/Z^0$      & $\alpha^5\,Z^2$                          & 0.003\,162\,903 & 0.000\,0\,(1) & 0.000\,002 \\
\multicolumn{1}{l}{Total $g-g_e$:} \\
\multicolumn{1}{l}{Theory, this work}     &                                          &  &  & -188.847\,(5) \\
%$g$          &                                          &  &  & 2002130.457\,(5) \\
\multicolumn{1}{l}{Theory \cite{glazov:04:pra}}  &                            &  &  & -188.819\,(19)
\end{tabular}
\end{ruledtabular}
\end{table*}

%\bibliographystyle{../bibtex/phaip30}
%\bibliography{../bibtex/hfst}

\end{document}